\documentclass[twocolumn]{aastex631}
\usepackage{natbib}
\usepackage{bm}
\begin{document}
\title{Non-thermal motions and atmospheric heating of cool stars}
\author{S. Boro Saikia}
\affiliation{University of Vienna, Department of Astrophysics,
              T\"urkenschanzstrasse 17, 1180 Vienna, Austria}

\author{ T. Lueftinger}
\affiliation{European Space Agency, European Space Research and Technology Centre, Keplerlaan 1, 2201-AZ Noordwijk, The Netherlands}
\author{V.~S. Airapetian}
\affiliation{NASA Goddard Space Flight Center, Greenbelt, MD USA}
\affiliation{American University, Washington, DC, USA}
\author{T. Ayres}
\affiliation{ Center for Astrophysics and Space Astronomy, 389 UCB, University of Colorado, Boulder, CO 80309, USA}
\author{M. Bartel}
\affiliation{University of Vienna, Department of Astrophysics,
              T\"urkenschanzstrasse 17, 1180 Vienna, Austria}

\author{M. Guedel}
\affiliation{University of Vienna, Department of Astrophysics,
              T\"urkenschanzstrasse 17, 1180 Vienna, Austria}

\author{M. Jin}
\affiliation{Lockheed Martin Solar and Astrophysics Lab (LMSAL), Palo Alto, CA 94304, USA}
\author{K. G. Kislyakova} 
\affiliation{University of Vienna, Department of Astrophysics,
              T\"urkenschanzstrasse 17, 1180 Vienna, Austria}
\author{P. Testa}
\affiliation{Harvard-Smithsonian Center for Astrophysics, 60 Garden St, Cambridge, MA 02193, USA}
\begin{abstract}
The magnetic processes 
associated with the non-thermal broadening of 
optically thin emission
lines {appear to} carry enough energy to 
heat the corona and accelerate the solar wind. We 
investigate whether non-thermal motions 
in cool stars exhibit the same behaviour
as on the Sun by analysing archival stellar 
spectra taken 
by the Hubble Space Telescope, and full-disc 
Solar spectra taken by the 
Interface Region Imaging Spectrograph. 
We determined the non-thermal velocities 
by measuring the excess broadening in optically
thin emission lines formed in the 
stellar atmosphere; the chromosphere, the 
transition region and the corona.
Assuming the non-thermal broadening is caused by 
the {presence} of Alfv\'en waves, we also
determined the associated wave energy densities.
Our results show that, with a non-thermal velocity of
$\sim$23 km~s$^{-1}$ the Sun-as-a-star results 
are in very good agreement with values
obtained from spatially-resolved solar observations.
The non-thermal broadening in our sample
show correlation to stellar rotation,
with the strength of the
non-thermal velocity decreasing with decreasing 
rotation rate. Finally, 
the non-thermal velocity in cool Sun-like stars
varies with atmospheric height or temperature
of the emission lines, and peaks
at transition region temperatures. This points towards
a solar-like Alfv\'en wave driven heating in stellar
atmospheres. However, the
peak is at a lower temperature in 
some cool stars suggesting that,
other magnetic process such as flaring events could also dominate.

\end{abstract}
\section{Introduction} \label{sec:intro}
One of the key drivers of atmospheric loss in (exo)planets 
orbiting cool main-sequence stars
is the magnetised winds of their central
star \citep{kislyakova14,airapetian17}. In cool stars like our Sun, the
wind {is driven by non-thermal 
processes in the stellar corona and interface region}
\footnote{includes the chromosphere and 
the transition region, which act as an interface between the 
photosphere and the corona}, which include
magnetic reconnection, propagation and 
{presence} of Alfv\'en waves, turbulence, 
flares and other explosive 
events \citep[e.g.,][and references therein]{mariska92,depontieu21}.
Out of these different processes 
Alfv\'en waves have emerged as the 
dominant heating mechanism in
solar and stellar wind models 
\citep{suzuki06, cranmer07,vanderholst14,lionello14,shoda19,reville2020}.
However, controversy still exists on the
contribution of these different
non-thermal processes towards coronal heating and wind
propagation \citep{cranmer19}.
Detailed comparative study
of the interface region in the Sun and other
cool stars can help shed light into these
different processes and constrain numerical
models of solar and stellar winds.

In the case of the Sun, far-ultraviolet (FUV) and 
extreme-ultraviolet (EUV) spectral lines provide important diagnostics 
of the plasma 
properties in the interface region. While flux ratios 
of certain spectral lines can be used to estimate the density
of the emitting plasma \citep{polito16,young18}, other diagnostics such as
Doppler shifts \citep{cheung15,testa16}, asymmetries
\citep{martinez-sykora11}, and non-thermal broadening \citep{chae98,depontieu15} 
of spectral lines provide valuable constraints 
to solar coronal heating and wind 
acceleration models. 
Here, we focus on the analysis of non-thermal broadening of FUV 
lines from cool stars, 
where the
non-thermal broadening is 
the excess broadening of a spectral line on top of the thermal and 
instrumental broadening. 

Instruments such as the
{ {\it Skylab} spectrograph \citep{tousey73,reeves76}, the 
Solar Ultraviolet Measurements of 
Emitted Radiation (SUMER) instrument aboard
the Solar and Heliophysics Observatory 
\citep[SOHO,][]{wilhelm95}, and the Interface Region
Imaging Spectrograph \citep[IRIS,][]{depontieu14}} have provided
new insights on the non-thermal 
broadening of interface region 
emission lines.
The non-thermal velocities, 
measured in the quiet Sun, range from 
5 to 30 km~s$^{-1}$, and show a correlation 
with the emission line temperature or the
atmospheric height
\citep{boland75,doschek76,mariska92,chae98,teriaca99}.
The measured non-thermal 
velocities increase with 
an increasing temperature, 
and reach a peak in the upper transition region, 
which is then followed by a decline towards 
higher
temperatures in the corona.
Correlations have been also reported 
between the non-thermal velocities and 
the line intensities, however
the correlations gets weaker with increase in 
temperature \citep{chae98}.
These observed properties of non-thermal velocities
{suggest that magnetic processes in the solar
atmosphere } drive
solar coronal heating and wind acceleration. 

While solar observations provide us with an unique opportunity to 
obtain an in-depth knowledge of the physical processes in a cool
main sequence star, it is a single data point 
amongst the hundreds of thousands of cool stars in 
our neighbourhood. To obtain a general 
understanding of non-thermal broadening in stellar
atmospheres and their connection to stellar fundamental 
properties such as mass and rotation, we must also
examine cool stars other than the Sun.
Measurements
of non-thermal velocities in cool stars are limited, and 
most of our knowledge of cool stellar non-thermal velocities
come from observations taken by 
the Hubble Space Telescope's Goddard High Resolution
Spectrograph
\citep[HST/GHRS,][and references therein]{linsky94,wood97}.
The interface region emission lines used in these
studies not only showed large non-thermal broadening,
but also a non-Gaussian profile with strong emission around
the wings. Although not as prominent as in stellar spectra, 
such a shape is also known to exist in solar
emission lines \citep{kjeldseth77,peter01,ayres21}.
To account for the broad wings the spectral lines 
are often modelled using a double Gaussian profile
with a narrow and a broad component. The non-thermal velocities 
are then calculated for both the narrow and the broad component
individually. 

The presence of the broad component adds an
additional layer of complexity in the interpretation
of the measured non-thermal broadening. 
In the case of giant and super-giant stars, 
anisotropically distributed turbulence along the 
line of sight has been proposed as 
a mechanism behind enhanced wings of 
non-thermally broadened lines 
\citep{robinson96,airapetian00}.
Based on observations of 
cool stars, 
\citet{wood97} concluded that the non-thermal broadening in 
the narrow component could be attributed to turbulence or
Alfv\'en waves, whereas the broad component is 
generated by micro-flares.
A Sun-as-a-star study 
by \citet{peter06}
provides an alternative theory, that the broad 
component is related to 
the underlying magnetic network in the chromosphere, 
and the non-thermal broadening of the narrow component
is a better indicative for coronal and wind heating.
A recent study of solar 
emission lines by \citet{ayres21} 
shows that the non-Gaussian nature of the lines prevail
even at the finest spatial scales, suggesting that 
the observed velocity distribution might be 
non-Maxwellian in nature. However, further investigations
are needed to explore the true nature of these 
emission lines. 

In this work we investigate the non-thermal
broadening in a sample of cool stars,
based on archival measurements of HST's 
Cosmics Origins Spectrograph (COS). 
We use a double Gaussian profile to
model the emission lines, and only consider the
narrow component of the Gaussian while
calculating the non-thermal 
broadening of the sample. 
We investigate the correlation 
between the measured non-thermal velocities and 
stellar properties at different formation
temperatures, and compare our results to
the non-thermal velocities measured using full-disk
solar data.  In 
section 2 we discuss the archival data set, followed by data
analysis in Section 3. The results are discussed in Section 4, 
and the conclusions in Section 5.

\startlongtable
\begin{deluxetable*}{lcccccccccc}
\tabletypesize{\footnotesize}
\tablewidth{0.99\textwidth}
\tablecaption{ \label{table1} Stellar properties \label{table1}}
\tablehead{
\colhead{HD name}             &other name  &\colhead{mass}              &\colhead{radius}              &\colhead{$T_\mathrm{eff}$}              &\colhead{$v_\mathrm{rad}$}
              &\colhead{$P_\mathrm{rot}$}               &\colhead{$\log {R'_\mathrm{HK}}$}              &\colhead{$\xi_\mathrm{Si1393}$}              &\colhead{planet host}              &\colhead{additional refs}
\\
&&$\mathrm{M_\odot}$ &$\mathrm{R_\odot}$&K&km~s$^{-1}$&days& &km~s$^{-1}$& &
}
\startdata
HD 75732              &55 Cnc       &0.92    &0.95    &5235      &27.8     &42.0             &-5.01      &    25.9   &Yes               &1, 2\\                                  
HD 129333             &EK Dra       & 0.89   &0.91    &5845      &-20.0    &2.6              &-4.10      &    38.0   &No               &1, 3\\                                 
HD 209458             &	HIP 108859  &1.10    &1.1     &6099      &-14.7    &11.4             &-4.88      &    25.4   &Yes               &4, 5\\                                  
HD 69830              &LHS 245      & 0.86   &0.90    &5361      &32.2     &41.2             &-5.02      &    24.9   &Yes               &1\\                                     
HD 154345             &LHS 3260     &0.90    &0.84    &5468      &-46.6    &31.0             &-4.80      &    23.9   &Yes                &1\\                                    
HD 204961             & GJ 832      &0.45    &0.48    &3472      &13.0     &40.0             &-5.07      &    14.8   &Yes               &1, 6\\                                  
HD 38529              &	HIP 27253   &..      &..      &5697      &30.0     &35.7             &..           &    35.2   &Yes                 &1\\                                   
..                    &LTT 2050     &0.48    &0.45    &3633      &-6.7     &3.7              &..           &    14.7   &No                 &7, 8\\                               
..                    &LP 415-1619  &..      &0.58    &3420      &55.0     &..               &..           &    26.3   &No                 &1, 9, 10\\                           
HD 121504             &LTT 5432     &1.18&..      &6075      &19.6     &8.6              &-4.75      &    29.7   &Yes               &1, 11\\                                 
..                    &GJ 3470      &0.45&0.42    &3600      &26.3     &20.7             &..           &    17.8   &Yes                 &12, 13\\                              
HD 3651               &54 Psc       & 0.80 &0.88    &5221      &-32.9    &37.0             &-5.11      &    22.4   &Yes               &1\\                                     
HD 38459              &LTT 2361     &0.88&..      &5320      &26.0     &12.0             &-4.46      &    17.1   &No               &14, 15\\                               
HD 97658              &GJ 3651      & 0.90&0.74    &5148      &-1.5     &38.5             &-5.04      &    20.4   &Yes               &1\\                                     
..                    &WASP-69      &0.83    &0.81    &4715      &-9.4     &23.1             &..           &    21.1   &Yes                 & 1,16\\                               
HD 172669             &	HIP 91210   &1.08    &1.00    &5964      &-6.7     &..               &..           &    17.2   &No                 &17, 18\\                             
HD 1461               &LTT 149      &1.14    &1.10    &5765      &-10.0    &29.0             &-5.33      &    28.9   &Yes               &1\\                                     
HD 95128              &47 UMa       &1.29    &1.22    &5882      &11.5     &24.0             &-4.90      &    29.3   &Yes                &1\\                                    
HD 285968             &GJ 176       &0.45    &0.45    &3416      &26.0     &38.9             &-4.76      &    15.9   &Yes               &1,19, 20\\                              
HD 62850              &	HIP 37563   &1.01    &1.05    &5700      &17.9     &..               &-4.27      &    31.4   &No               &21, 22, 23, 24\\         
HD 22049              &$\epsilon$ Eri  &0.71    &0.72    &5146      &16.0     &11.7             &-4.50      &    23.2   &Yes                &1\\                                    
HD 13445              &GJ 86    &0.93    &0.80    &5151      &55.0     &30.0             &..           &    19.4   &Yes                 &1\\                                   
HD 7924               &HIP 6379 &0.80    &0.75    &5177      &-22.0    &38.0             &-4.90      &    18.2   &Yes                &1\\                                    
..                    &HAT-P-11    &0.81    &0.75    &4780      &-63.7    &30.5             &-4.65      &    18.7   &Yes               &1, 25\\                                 
HD 189733             &GJ 4130&0.82    &0.76    &5050      &-2.3     &13.4             &-4.50      &    21.8   &Yes                &1, 26\\                                
HD 156384C            &GJ 667 C&0.33    &..      &3600      &6.0      &105.0            &-5.30      &    10.4   &Yes                &1, 27, 28, 29 \\                       
HD 40307              &	GJ 2046&0.78    &0.71    &4827      &31.3     &48.0             &-5.25      &    20.1   &Yes               &1\\                                     
..                    &GJ 849  &0.49    &0.46    &3601      &-32.0    &39.2             &-4.95      &    15.7   &Yes               & 30, 31, 32 \\                          
HD 146233             &18 Sco& 0.98   &1.02    &5791      &10.0     &..               &-4.90      &    24.8   &No                &\\                                    
HD 104067             &LTT 4461& 0.91   &0.75    &4956      &15.0     &34.7             &-4.80      &    18.4   &Yes                &1\\                                    
HD 128311             &	GJ 3860& 1.32   &0.73    &4965      &-9.0     &14.0             &-4.49      &    23.6   &Yes               &1\\                                     
HD 136118             &	HIP 74948& 1.25   &1.74    &6097      &-3.0     &12.2             &-4.79      &    23.8   &No               &1\\                                    
HD 201091             &61 Cyg A&0.66    &0.67    &4545      &-65.5    &35.0             &-4.70      &    19.8   &No                & 33, 34 \\                            
HD 41004              &	HIP 28393&0.70    &..      &5010      &42.5     &27.0             &-4.67      &    23.6   &Yes               &35\\                                    
HD 99492              &83 Leo B& 1.24   &0.76    &4955      &3.0      &45.0             &-5.00      &    17.7   &Yes                &1\\                                    
HD 10647              &q1 Eri& 0.98   &1.10    &6105      &27.6     &10.0             &..           &    26.0   &Yes                 &1\\                                   
..                    &GJ 876  &0.37    &0.38    &3130      &-1.5     &81.0             &-5.00      &    13.6   &Yes                & 36, 37\\                              
HD 92788              &LTT 3928& 1.64   &0.99    &5836      &-4.0     &31.7             &-4.95      &    17.9   &Yes               &1\\                                     
HD 150706             &GJ 632&1.17    &0.96    &5961      &-17.0    &..               &-4.40      &    25.1   &Yes                &38\\                                   
HD 39091              &$\pi$ Men&1.12    &1.15    &5950      &10.0     &33.9             &-4.89      &    30.3   &Yes               &1\\                                     
HD 85512              &LHS 2201&0.69    &..      &4715      &-10.0    &47.1             &-4.96      &    17.0   &Yes               &1, 39\\                                 
HD 192263             &	HIP 99711& 0.83   &0.75    &4975      &-10.0    &24.5             &-4.57      &    18.0   &Yes               &1\\                                     
..                    &GJ 581 &0.31    &0.29    &3498      &-16.0    &94.2             &-5.43      &    9.9   &Yes               &1, 40, 41\\                             
HD 143761             &$\rho$ CrB            & 1.47   &1.32    &5823      &18.0     &18.5             &-4.95      &    18.0   &Yes               &1\\                                     
HD 186427             &16 Cyg B& 1.10   &1.15    &5674      &-28.0    &29.1             &-4.90      &    27.1   &Yes                &1\\                                    
HD 72905              &$\pi^1$ UMa&1.00    &0.96    &5873      &-12.0    &5.0              &-4.30      &    27.0   &No                & 42, 43, 44\\                         
HD 161897             &	HIP 86540& 1.01   &0.86    &5623      &-16.5    &..               &-4.77      &    17.7   &No               &\\                                     
HD 192310             &LHS 488& 0.84   &0.81    &5080      &-54.0    &47.7             &-5.30      &    16.8   &Yes                &1\\                                    
HD 24636              &HIP 17764&1.39    &1.37    &6831      &14.5     &..               &..           &    57.6   &No                 &17, 23, 45 \\                        
HD 25825              &LP 15-582& 1.01   &1.08    &5941      &37.0     &..               &-4.34      &    23.3   &No               &\\                                     
HD 160691             &$\mu$ Ara&1.08    &..      &5813      &-12.0    &31.0             &-4.97      &    22.0   &Yes              &46,47\\                     
HD 39587              &$\chi^1$ Ori& 0.82   &1.01    &5882      &-13.4    &5.2              &-4.40      &    30.7   &No               &\\                   
HD 186408             &16 Cyg A& 1.25   &1.25    &5781      &-27.5    &26.9             &-4.98      &    25.5   &No                &1\\                  
HD 197037             &LTT 16037 &1.11    &1.15    &6150      &8.0      &19.1                       &..&    18.8   &Yes               &1,48\\                    
HD 1835               &9 Cet& 0.98   &0.96    &5837      &-2.5     &..               &-4.40      &    29   &No                &\\     
\enddata
\tablecomments{Columns 1 to 9: HD name, other name, mass, radius, effective temperature, radial velocity, rotation, stellar activity 
or $\log {R'_\mathrm{HK}}$, mean non-thermal velocity determined in this work from the \ion{Si}{4} 1393 \AA~line. 
The stellar parameters are taken from \citet{valenti05} or the following references,
\citet{france18}$^1$; \citet{poppenhaeger10}$^2$; \citet{wright11}$^3$;\citet{silvavalio08}$^4$;
\citet{borosaikia18}$^5$; \citet{wittenmyer14}$^6$; \citet{schweitzer19}$^7$; \citet{passegger20}$^8$;
\citet{kopytova16}$^9$; \citet{terrien15}$^{10}$; \citet{mayor04}$^{11}$; \citet{bonfils12}$^{12}$;
\citet{palle20}$^{13}$; \citet{pavlenko19}$^{14}$; \citet{hojjatpanah20}$^{15}$; \citet{anderson14}$^{16}$;
\citet{mcdonald12}$^{17}$; \citet{kervella19}$^{18}$; \citet{peacock19}$^{19}$; \citet{endl08}$^{20}$;
\citet{muirhead18}$^{21}$; \citet{cutispoto02}$^{22}$; \citet{bochanski18}$^{23}$; \citet{nielsen19}$^{24}$;
\citet{bakos10}$^{25}$; \citet{bouchy05}$^{26}$; \citet{anglada12}$^{27}$; \citet{anglada13}$^{28}$;
\citet{delfosse13}$^{29}$; \citet{schoefer19}$^{30}$; \citet{butler06}$^{31}$; \citet{feng15}$^{32}$;
\citet{borosaikia16}$^{33}$; \citet{kervella08}$^{34}$; \citet{santos02}$^{35}$; \citet{vonbraun14}$^{36}$;
\citet{youngblood17}$^{37}$; \citet{boisse12}$^{38}$; \citet{pepe11}$^{39}$; \citet{bonfils05}$^{40}$;
\citet{vonbraun11}$^{41}$; \citet{cenarro07}$^{42}$; \citet{rosen16}$^{43}$; \citet{kochukhov20}$^{44}$;
\citet{gaspar16}$^{45}$; \citet{butler01}$^{46}$; \citet{santos04}$^{47}$; \citet{robertson12}$^{48}$
}
\end{deluxetable*}

 \section{Observations and data analysis}
The stellar data used in this work were obtained from 
the Mikulski Archive for Space Telescopes (MAST) at the Space Telescope Science Institute. 
The specific observations analyzed can be accessed via \dataset[DOI: 10.17909/wvzj-wd79]{https://doi.org/10.17909/wvzj-wd79}.
Table \ref{table1} 
lists the properties of our stellar sample of
55 stars, all of which 
were observed by HST COS 
\citep{froning09}, which is a highly-sensitive 
moderate resolution spectrograph with a spectral resolution of
1600-24000 over a wavelength range of 1150-3200~\AA. 
We also analysed
solar spectra taken by IRIS \citep{depontieu14},
which is a NASA Small Explorer satellite that
takes high-resolution UV images and spectra
of the Sun; spectral resolution of 
$\sim$50,000. 
In addition to daily raster scans
of the Sun IRIS provides 
monthly full-disc mosaics
in 6 spectral windows, enabling us to compare Sun-as-a-star
spectral lines to the stellar sample.

\subsection{Interface region emission lines}\label{lines}
Non-thermal velocities in stellar interface regions can 
be determined from optically thin
emission lines. In this work we analysed 
optically thin FUV emission lines formed 
in the interface layer; the 
chromosphere and the transition region. 
Additionally, we also included a coronal emission 
line in our analysis. 
The line selection was 
based on the 
wavelength coverage of COS and IRIS, and on the
signal-to-noise ratio of the archival observations. 
Figures~\ref{fig1}~and~\ref{fig2} show the emission 
lines analysed in this work (absolute flux densities
are shown in Figure~\ref{appendix_fig1}). 

\subsubsection{Chromospheric emission lines}
The most well-studied
chromospheric lines are optically thick, 
which include the \ion{Mg}{2} h and k lines 
at 2803.52 \AA~and 2796.34 \AA, 
and the \ion{Ca}{2} h and k lines at 3936.85 \AA~
and 3933.64 \AA,~respectively. In
recent years, due to its
optically thin nature, the 
\ion{O}{1} 1355.6 \AA~line has emerged as a strong
diagnostics for non-thermal motions in 
the chromosphere.
According to
\citet{lin15}, 
the \ion{O}{1} line forms at
a wide range of heights 
at the middle of the chromosphere,
and in this work we adopt a 
temperature of 20,000 K \citep{teriaca99}.
Due to its 
formation in the chromosphere
the \ion{O}{1} line is a very good diagnostic 
tool not only for non-thermal broadening but
also as a proxy for magnetic activity in 
the stellar chromosphere. 
However, careful analysis is required as the \ion{O}{1} line 
blends with the \ion{C}{1} line at 1355.8 \AA,~
as shown in Figures~\ref{fig2}~and~\ref{appendix_fig1}.

\subsubsection{Transition region emission lines}
The transition region is the second layer
of the interface region, where the temperature
rises from a few thousands of Kelvin to a million
degree Kelvin. It is also the region where 
multiple different optically thin emission lines 
form, providing an excellent diagnostic for plasma
motions. We included five transition region emission
lines in our study, which include the  
\ion{Si}{4} doublet at 1393.76~\AA~and 1402.77 \AA, 
the \ion{C}{4} doublet at 1548.195~\AA~and 1550.77 \AA,
and \ion{O}{4} at 1401.156 \AA,~respectively.
Figure~\ref{appendix_fig1} shows these five 
transition region emission lines for the young exoplanet
host star $\epsilon$ Eri.

The \ion{Si}{4} doublet lines are resonance lines formed under
optically thin collisionally excited conditions with a peak formation 
temperature of $\sim$80,000 K \citep{peter14} and lie
at the middle of the transition region. The \ion{Si}{4} line formed at
1393 \AA~(decimals are ignored for simplicity in the 
rest of the text) has a stronger intensity than the line
formed at 1402 \AA. Under optically thin conditions
the flux ratio of these two lines (1393/1402) is 2, while 
under the optically thick conditions the ratio is 
$\neq 2$. The \ion{Si}{4} doublet lines 
are the only lines for which we 
have included full-disc solar observations from IRIS.

\begin{figure}
\centering
\includegraphics[width=0.47\textwidth]{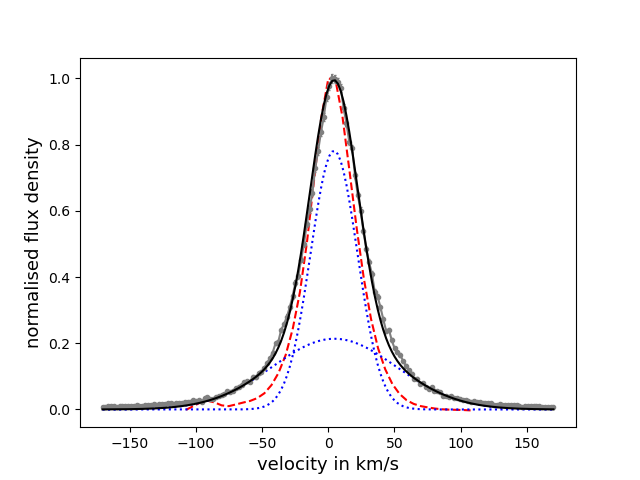}
\caption{HST COS \ion{Si}{4} 1393 \AA~ line of the active young
K dwarf $\epsilon$ Eri in grey. The best fit
Gaussian model is shown in black, followed by the narrow
and the broad component of the Gaussian fit in blue. 
A full-disc solar \ion{Si}{4}
1393 \AA~ line is shown for comparison in red. }
\label{fig1}
\end{figure}

The \ion{C}{4} doublet lines are resonance lines with 
a peak formation temperature at 100,000 K  \citep{teriaca99}.
The 1548 \AA~line originates from an atomic transition from
the ground to the 3rd energy level and the 1550 \AA~line is formed
due to a transition from the ground to the 2nd energy level.
The \ion{C}{4} doublet have the strongest intensity when
compared to the other emission lines used in this work.
Under non-flaring solar conditions the flux ratio of the 
two lines in the doublet are shown to be $\sim$ 2,
suggesting optically thin conditions \citep{dere93}. 
Due to limited wavelength coverage the 
\ion{C}{4} line could be analysed for 
only one star in the sample, $\epsilon$ Eri.

\begin{figure*}
\centering
\includegraphics[width=1\textwidth]{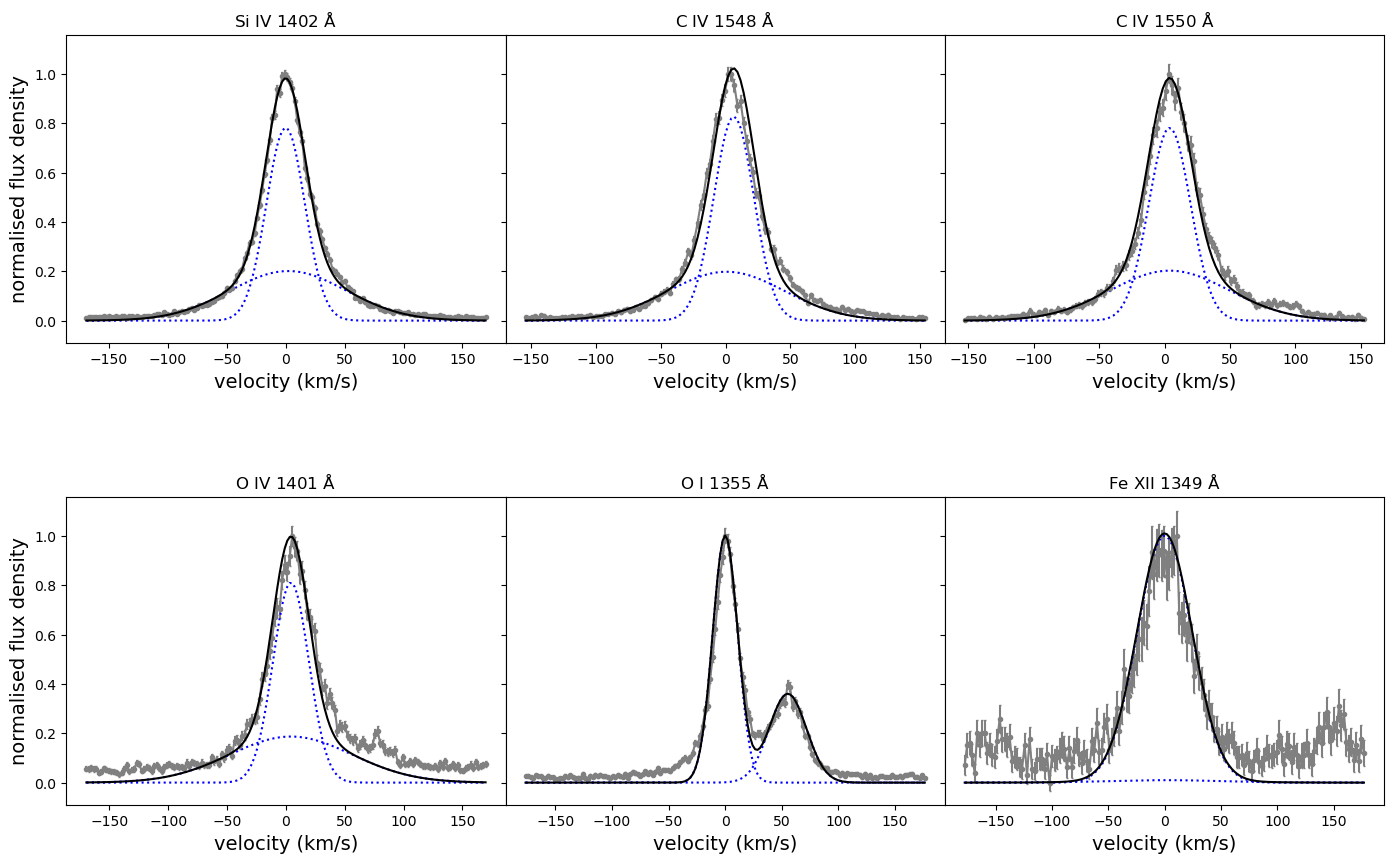}
\caption{Interface region and coronal emission lines of $\epsilon$
Eri in grey. The best fit Gaussian model is shown
in black followed by the narrow and the broad component
in blue.}
\label{fig2}
\end{figure*}

The \ion{O}{4} line at 1401 \AA~is one of the five 
\ion{O}{4} inter-combination lines found close to the
\ion{Si}{4} line. With a formation temperature of 
150,000 K under equilibrium conditions \citep{depontieu14}
the \ion{O}{4} line is an important density diagnostic tool
\citep{keenan02}. The ratio of this line with the 
\ion{O}{4} line at 1399 \AA~ is widely used as
a density diagnostic tool for medium 
density plasma \citep{polito16}. In the Sun, 
the wings of \ion{O}{4} line are known
to be blended with the 
photospheric cool transition line 
of \ion{S}{1} \citep{polito16}.

\subsubsection{Coronal emission lines}
While the corona mostly emits
in X-rays, it can be also probed using a
limited number of FUV and EUV lines. The coronal
emission line included in this study is 
the \ion{Fe}{12} 1349.4 \AA~forbidden line with
a peak formation temperature of $\sim$ 1.5 MK.
Figure~\ref{appendix_fig1} shows the \ion{Fe}{12} line
of $\epsilon$ Eri, which has the 
poorest signal-to-noise ratio amongst
all the lines included in this work. 
Furthermore, the weak \ion{Fe}{12} line is only 
observable for two stars in our sample. 
Even in the Sun long exposure
times are required to obtain a good 
signal-to-noise ratio for this line \citep{testa16}.

Our results in Section \ref{results} are primarily based on the 
\ion{Si}{4} line at 1393 \AA,~ as it is most easily
detectable line for all stars included in our study. 
The results related to the 
other six emission lines are only discussed for a handful
of stars with a good signal-to-noise ratio.
\subsection{Gaussian model line fitting and non-thermal velocities}
In order to determine the non-thermal broadening
we first perform a Gaussian fit to the observed emission lines. 
The full width at half maximum (FWHM) of the  best fit model is then
used to measure the non-thermal broadening after subtracting the 
corresponding thermal broadening.

As shown in Figures \ref{fig1} and \ref{fig2}, the 
\ion{S}{4}, \ion{C}{4} and \ion{O}{4} lines were
fitted using a double Gaussian model, 
which includes a narrow component (NC) and a broad (BC) component. 
A slightly different Gaussian model was applied to 
the \ion{O}{1} and \ion{Fe}{12} lines.
The nearby \ion{C}{1} was included in the fitting 
procedure of the \ion{O}{1} line. The best fit model doesn't 
account for the broad wings of \ion{O}{1} suggesting 
an additional Gaussian component might be necessary,
which is not included here. 
Finally, a single Gaussian model is used to fit the 
\ion{Fe}{12} line. Analysis of IRIS \ion{Fe}{12} lines 
in the solar active regions by \citet{testa16} 
show that a single Gaussian model is sufficient to 
model this line. Figure~\ref{fig2} also suggests that the 
\ion{Fe}{12} line can be modelled using a single Gaussian model.
However it has been shown that, while  spectra with poor
signal-to-noise can be easily fit with a single Gaussian 
line profile, the same spectra might need a double Gaussian
model if the signal-to-noise ratio improves \citep{peter06}.
Finally, although a double Gaussian model is
applied to the \ion{O}{4} line, the 
\ion{S}{1} line blend is not included in the line
fitting process. Hence, care should 
be taken while interpreting the non-thermal
broadening determined using the 
\ion{O}{1}, \ion{Fe}{12}, and \ion{O}{4} lines.

Since our emission line profiles are composed of 
either a single or a double Gaussian, the 
FWHM of the best fit 
Gaussian models can be expressed as a combination of
thermal and non-thermal motions,

\begin{equation}
FWHM = \sqrt{
    4 ln 2 {\left (\frac{\lambda}{c}\right)}^2\left(\frac{2k_\mathrm{B}T}{M}+
    {\xi}^2 \right)}
\label{eq1}
\end{equation}  

where $\lambda$ is the rest wavelength of the emission line in \AA,
$c$ is the speed of light in km~s$^{-1}$, 
$k_\mathrm{B}$ is the Boltzmann constant, $T$ 
is the temperature of the plasma 
in Kelvin, $M$ is the mass of the ion
emitting the line in units of the hydrogen 
atom, and $\xi$ is the non-thermal velocity along
the line of sight in km~s$^{-1}$.

In addition to thermal and 
non-thermal broadening, other key
broadening mechanisms in the observed
emission lines are instrumental 
and rotational broadening. 
For rapidly rotating stars, especially
for chromospheric lines where the lines are
narrower than their transition region counterparts,
the affect of rotational broadening
should be included in the model. 
The rotational broadening for 
moderate to slowly rotating stars,
however,
is negligible compared 
to the other broadening mechanisms. 
The stellar sample included here mostly 
consists of moderate and slow rotators, 
hence rotational broadening 
is not included in our
current model.
To account for the instrumental broadening of COS
spectra, our Gaussian
models are convolved with the COS line spread function (LSF) before 
applying the fitting algorithm. 
The instrumental broadening of 
IRIS is reported to be $\sim$5.5 km~s$^{-1}$
\citep{depontieu14}, which is accounted for
in our calculation by including the 
$\Delta_\mathrm{inst}$ term in equation \ref{eq1},
as shown below,
\begin{equation}
FWHM = \sqrt{\Delta_\mathrm{inst}^2 +
    4 ln 2 {\left (\frac{\lambda}{c}\right)}^2\left(\frac{2k_\mathrm{B}T}{M}+
    {\xi}^2 \right)}
\label{eq2}
\end{equation}

Our fitting algorithm is based on 
{\sc python}, and is optimised by 
{\sc scipy's} least-squares minimisation. 
Additionally, we determine 
the goodness of our fits by calculating the $\chi^2$. 
To provide an error estimate we carried out multiple
Gaussian fits by including
a wide range of initial conditions to the
fitting algorithm. The standard deviation 
of the non-thermal 
velocities calculated from 
these Gaussian fits is 
taken as 
the dispersion, with an average 
dispersion of $\sim$2 km~s$^{-1}$.
Figures \ref{fig1} and \ref{fig2} show examples of 
the best-fit Gaussian models to $\epsilon$ Eri's emission lines 
observed by HST COS, where the line widths are much 
broader than what is expected from thermal broadening 
alone. As an example, the width of the \ion{Si}{4}
line due to thermal motions is $\sim$12 km~s$^{-1}$,
which is much smaller than the line width seen in 
Figure~\ref{fig1}. The average non-thermal velocity obtained for
this star after subtracting the thermal broadening is
23.2 km~s$^{-1}$ (Table~\ref{table1}). This clearly shows that
non-thermal broadening mechanisms dominate in the interface
region lines discussed here.

\subsection{Non-thermal energy carried by Alfv\'en waves}\label{energy2.4}
The observed non-thermal 
velocity could be attributed to multiple
different processes, out of which Alfv\'en waves 
have widely emerged
as the mechanism of choice 
for current coronal and wind 
heating models \citep{vanderholst14,lionello14,reville2020}.
If the excess energy required to
heat solar/stellar winds is provided 
by the {presence} of 
transverse Alfv\'en waves, 
then the associated wave energy 
density $w$ can be expressed as, 

\begin{equation}
w=\rho\delta{v^2}
\label{wave_eq}
\end{equation} 

where $\rho$ is the 
mass density and 
$\delta{v^2}$ is the 
wave velocity perturbation,
and both of these terms can be 
observationally constrained
using the emission 
lines discussed here. 
The wave velocity perturbation
$\delta{v^2}$ can be determined
directly from the 
observed non-thermal velocities 
$\xi^2=\frac{1}{2}\delta{v^2}$ 
\citep{banerjee98}. 
The mass density $\rho$ can
be determined from the electron number 
density $N_\mathrm{e}$, 
$\rho=m_\mathrm{H} N_\mathrm{e}$,
where $m_\mathrm{H}$ is the mass of a single 
hydrogen atom. Flux ratios
of certain interface region emission
lines act as a good diagnostic of the 
number density 
$N_\mathrm{e}$
\citep{keenan02,polito16}. However, it should be 
noted that this diagnostic method is sensitive 
to line formation temperature. As the temperature
in the interface region vary rapidly the density 
diagnostic will be strongly dependent on the 
line used. In this work
the \ion{O}{4} 1399/1401 line ratio 
was analysed to constrain $N_\mathrm{e}$ and $\rho$, 
as discussed in Section \ref{heating}. 
\section{Results and discussion} \label{results}
\subsection{Emission line shapes}
Figure~\ref{fig1} shows an example
\ion{Si}{4} 1393 \AA~line for the 
active young Sun $\epsilon$ Eri. 
The best fit double Gaussian model is
also shown, together with a full-disk
spectra of the Sun taken by the IRIS 
spectrograph. 
In the case of $\epsilon$ Eri 
strong excess emission close to the wings 
is detected in both the red and the blue
part of the spectral line. Such strong 
emission around the wings have been detected
in all of the stars in our sample.
In comparison the excess emission in
the full-disc solar spectral line
is relatively weaker, as shown
in Figure~\ref{fig1}.

In addition to the double Gaussian shape,
some of the emission lines exhibit distortions 
in the line profile, and in some cases shifts
of the central wavelength of the line profile
were also detected. In Fig.~\ref{fig2} the 
core of the \ion{C}{4} 1548 \AA~shows a distorted
profile shape. In the solar case 
distortions and variations in the
FUV emission lines are caused due 
to non-thermal motions 
along the line-of-sight \citep{phillips08}. 
Recently, such distortions were also
detected for the exoplanet host star
55 Cnc \citep{bourrier18}, where the
distortions and the reduction in flux in FUV
lines coincided with one of its planet, 
55 Cnc e's transit.
According to the authors the variation seen 
in these lines could not be explained by
stellar activity alone and could include 
contributions from 
possible star-planet interaction. 
A detailed analysis of these
line profile distortions and Doppler
shifts in stars, with and without exoplanets, 
has the potential to uncover a possible diagnostics
for star-planet-interaction, which 
is however beyond the scope of 
this work. 

\begin{figure}
\centering
\includegraphics[width=0.5\textwidth]{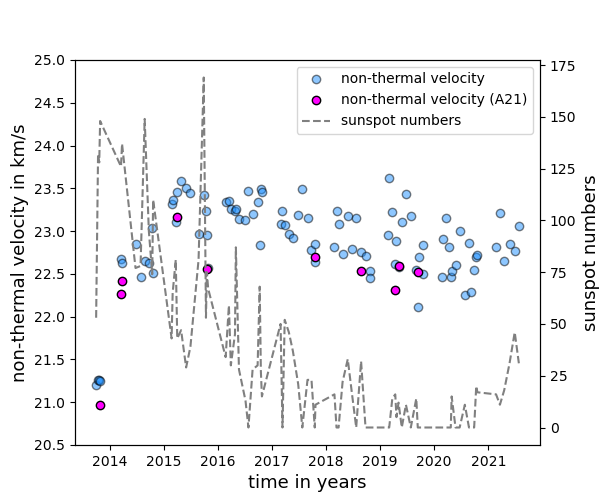}
\caption{Full-disc solar non-thermal velocities as a function of time.
The blue circles represent the 98 full-disc IRIS observations. The magenta circles
represent non-thermal velocities determined for the observations in A21. The dashed grey curve 
follows the sunspot cycle. The sunspot numbers were taken from World Data Centre SILSO, 
Royal Observatory of Belgium, Brussels.}
\label{fig4}
\end{figure}
\subsection{Non-thermal velocity at the solar transition region}
To determine the non-thermal velocity 
in the solar atmosphere we analysed 
98 full-disc
mosaics surrounding the \ion{Si}{4} 1393 \AA~line
taken by the IRIS mission,
which cover a period  
from the declining phase of 
solar cycle 24 to the increasing 
phase of cycle 25 (September 2013 
to June 2022). 
The non-thermal velocities determined
from these observations have a 
mean of 22.9 km~s$^{-1}$ and a
standard deviation of 0.5 km~s$^{-1}$,
which are in strong agreement with 
measurements 
of the Quiet Sun (QS) by \citet{dere93}
and \citet{chae98}. 
The non-thermal velocities
determined from full-disk IRIS
mosaics also agree with 
results from spatially resolved IRIS 
observations of the \ion{Si}{4}
1402 \AA~line, where the non-thermal
velocity was determined to be $\sim$20 km~s$^{-1}$
\citep{depontieu15}.

Ten out of the 98 full-disc
IRIS mosaics used here were also 
analysed
by \citet{ayres21} as part of their
Sun-as-a-star study (referred to
as A21 from here on), where 
they included 10 full-disc mosaics taken between October 
2013 and September 2019. The data analysed 
by A21 
were corrected for cosmic ray hits 
and missing data, and for this data-set we 
obtained a 
mean non-thermal velocity of 22.4 km~s$^{-1}$ 
with a
standard deviation of 0.6 km~s$^{-1}$,
which is in agreement with the results
obtained from the full IRIS sample of 98 
observations. For a detailed comparison 
between individual observations please
see Appendix ~\ref{comparison}.

\begin{figure*}
\centering
\includegraphics[width=0.5\textwidth]
{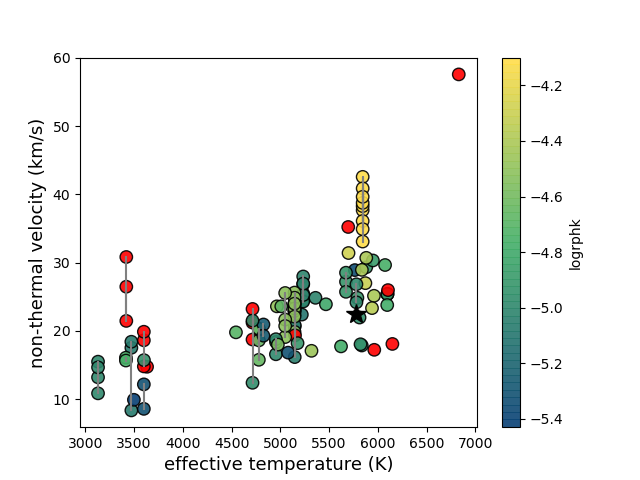}~~
\includegraphics[width=0.5\textwidth]
{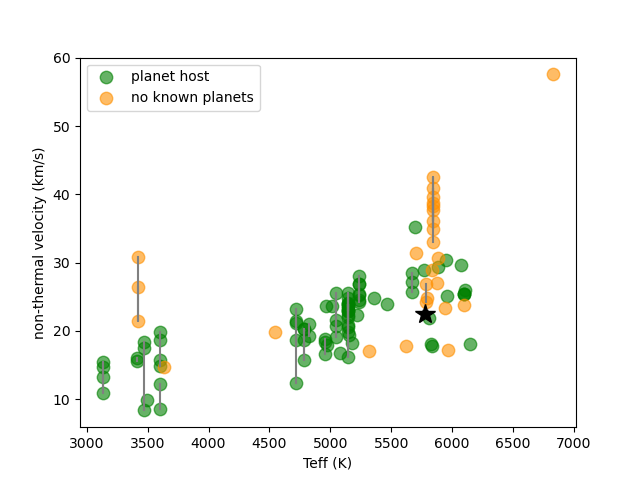}
\caption{Non-thermal velocity, derived from the \ion{Si}{4} 1393~\AA~line width, as a function of stellar
effective temperature. {\it Left}: The colour bar represents
stellar chromospheric activity 
($\log R'_\mathrm{HK}$) taken from the literature. 
Stars without known literature $\log R'_\mathrm{HK}$ values
are marked in red. {\it Right}: Same as the Figure on the left 
except green marks the known exoplanet hosts and orange represents
stars without known exoplanets. }
\label{fig5}
\end{figure*}

Since IRIS observations cover the decreasing 
phase of solar cycle 24 and the increasing phase
of cycle 25, we also investigate any possible correlation
between full-disc solar 
non-thermal velocities and magnetic spot emergence.
Figure~\ref{fig4} shows the evolution of the 
non-thermal velocity with the solar cycle. 
There is a phase 
delay between the peaks of the two measurements
with the non-thermal velocity peaking at least a year later and
showing a general flat distribution towards the declining phase
of cycle 24.
The lower envelope of the
non-thermal velocities for the entire sample of 98 observations 
exhibit {a similar trend as}
the sunspot numbers as they progress from cycle 24 to 25. 
The non-thermal velocities determined for the observations
taken from A21 agree well with the full IRIS sample used here. 
Sunspot numbers are 
a proxy for photospheric magnetic activity and 
non-thermal velocities are diagnostics for 
magnetic processes in the interface region. Since
the interface region magnetic field has roots
in the photosphere, it is not surprising
to see a general agreement in Figure~\ref{fig4} despite 
the difference in spatial scales. 

\subsection{Non-thermal velocity vs stellar properties}
We determine the non-thermal velocities for the stars 
listed in Table~\ref{table1} from the \ion{Si}{4} 1393~\AA~line
profile, and 
investigate their relationship with stellar
properties such as rotation and effective temperature.

The spectral coverage of our sample lies between late
F and mid M dwarfs, with G and K dwarfs being the dominant
majority, as shown in Figure~\ref{fig5}. The
chromospheric activity of the sample indicate
mostly intermediate to low activity stars, with a few
active stars in the mix.
The solar non-thermal velocity, determined
in the previous section, is close to the average
non-thermal velocity for its spectral type. 

The non-thermal velocities 
show a weak dependence on the stellar spectral type, with 
an overall decrease in velocity towards late type
dwarfs, as shown in 
Figure~\ref{fig5}. 
{M dwarfs
are known to have stronger surface
magnetic fields when compared to G and K
dwarfs \citep{shulyak17}. However, 
late-type dwarfs have smaller surface 
convective velocities
compared to G dwarfs, which corresponds to smaller 
amplitudes for Alfv\'en waves  \citep{sakaue21} 
and non-thermal velocities. This could explain the weak 
dependence of non-thermal velocity on spectral type.
Additionally, as shown in the right panel of 
Figure~\ref{fig5}, 
the majority of the 
stars in our sample are exoplanet 
hosts. Our sample is biased
towards inactive cool stars and 
only includes partially convective
M dwarfs. A majority
of the known partially convective M dwarfs
outside of young associations are inactive
and believed to be older, which could also
contribute towards the relatively low 
non-thermal broadening in M dwarf spectra
seen in this work.
This decreasing trend towards early 
M dwarfs has been also reported in
multiple chromospheric activity studies of 
late-type dwarfs \citep{reiners12,astudillo17,borosaikia18}.
Hence, rotation or age should be also 
included when investigating the non-thermal
broadening-stellar property relation for
our sample. }

\begin{figure*}
\centering
\includegraphics[width=0.5\textwidth]
{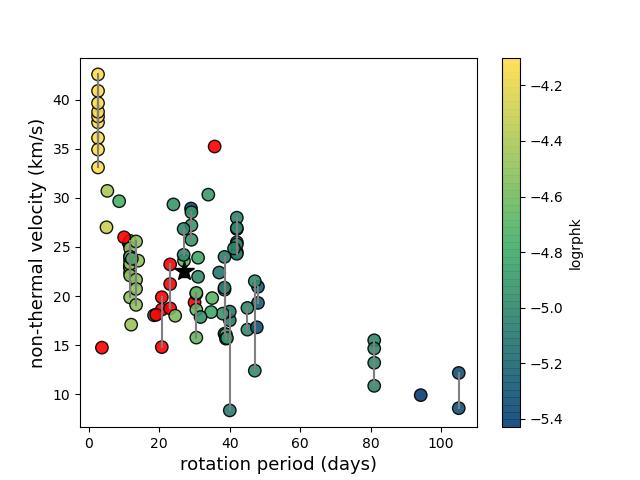}~~
\includegraphics[width=0.5\textwidth]
{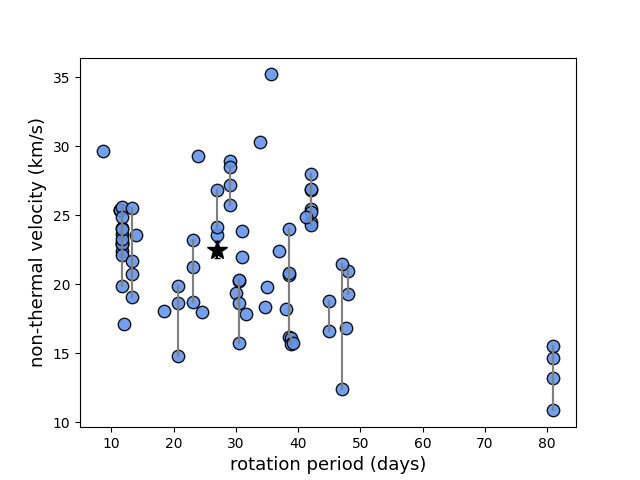}\\
\caption{ {\it Left:} Non-thermal velocity as a function of stellar rotation
period. Colour scale same as the left panel of Figure~\ref{fig5}. {\it Right:} Same as the figure 
on the left but only stars with $v \sin i<$5 km~s$^{-1}$ are included.}
\label{fig6}
\end{figure*}

Stellar non-thermal
velocity shows a clear dependence on
rotation, as presented in Figure~\ref{fig6}.
The late-type stars in Figure~\ref{fig5} 
with the lowest non-thermal velocity 
are indeed older slowly rotating dwarfs.
Overall the non-thermal 
velocity shows a linear
dependence on rotation, but within the 
intermediate to rapidly rotating cohort 
the non-thermal velocity exhibits
a flat distribution between 15-30 km~s$^{-1}$.
The rapidly rotating young Sun EK Dra
with non-thermal velocity in the range
of 30-40 km~s$^{-1}$ gives the impression 
of an increase with rapid rotation {in the 
left panel of Figure~\ref{fig6}}. However,
at a rotation period of $\sim$2 days 
rotational effects could distort 
and broaden the observed
spectral line. Furthermore, some of the EK Dra 
observations are affected by a flaring event 
\citep{ayres10, ayres15}. Hence, the 
quiescent non-thermal
velocity of EK Dra could be much lower than 
reported here. 

{Since the influence of
rotational broadening is not included 
in the 
model spectra the stellar sample
was additionally reduced to include only slowly
rotating stars. The right panel
of Figure~\ref{fig6} shows the non-thermal velocity
of a smaller sample of stars with low rotational velocity, 
$v\sin i <5$ km~s$^{-1}$ \footnote{Five stars, including 
two M dwarfs, didn't have any $v\sin i$ measurements so those
stars were also omitted in the right panel of 
Figure~\ref{fig6}.}.
The overall trend is the same as the left
panel of Figure~\ref{fig6}. The non-thermal 
velocity of stars similar to the Sun in 
rotation period lie in the range of 15-30 km~s$^{-1}$, and 
the single slowly rotating M dwarf exhibits a non-thermal 
velocity in the range of 10-15 km~s$^{-1}$. 
Future observations of both
rapidly and slowly rotating stars, together with a more complex
model, would help to further 
constrain the relationship between non-thermal
velocity and rotation in 
Figure~\ref{fig6}.}

\subsection{Non-thermal broadening vs \ion{Si}{4} 1393 flux}
The \ion{Si}{4} 1393~\AA~line is also considered 
to be an indirect tracer of
stellar magnetic field, and a proxy for 
stellar EUV activity \citep{france18}. 
Hence, we investigate possible
correlations between the 
\ion{Si}{4} 1393 flux and the 
measured non-thermal velocities. 
The flux in the \ion{Si}{4} 1393
line, $F_\mathrm{SiIV1393}$, is calculated by 
performing an integration centred around the line.
Since our sample consists of stars of different 
spectral types we normalise the measured flux by 
the bolometric flux $F_\mathrm{bol}$.
\begin{equation}
F_\mathrm{bol}=\sigma T_\mathrm{eff}^4 \left(\frac{R}{d}\right)^2
\end{equation}

where $\sigma$ is the Stefan-Boltzmann constant, 
$T_\mathrm{eff}$ is the stellar effective temperature, 
$R$ is the stellar radius, and $d$ is the distance. 

\begin{figure}
\centering
\includegraphics[width=0.5\textwidth]{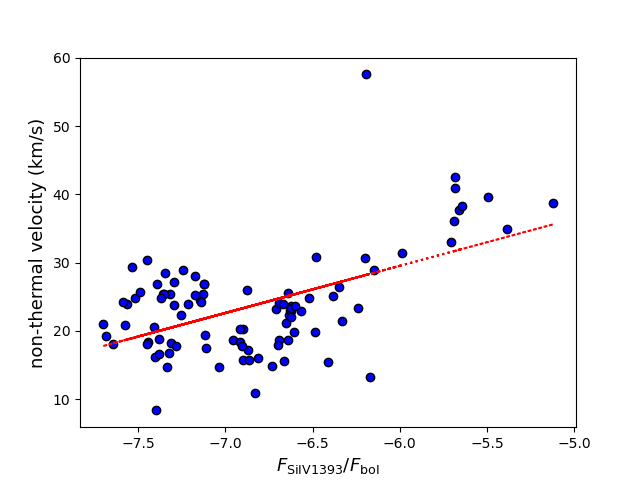}
\caption{Non-thermal velocity vs \ion{Si}{4} 1393 flux. The dotted line
is a linear fit to the data, where the R$^2$ is 0.3.}
\label{fig8}
\end{figure}

Figure~\ref{fig8} shows the 
\ion{S}{4} 1393 \AA~non-thermal velocity
as a function of 
the normalised \ion{Si}{4} flux ${F_\mathrm{SiIV1393}}/{F_\mathrm{bol}}$. 
The non-thermal velocities exhibit
a weak correlation with the measured flux, 
with a coefficient of determination ($\mathrm{R^2}$) of 0.3.  
The correlation appears to be 
stronger for ${F_\mathrm{SiIV1393}}/{F_\mathrm{bol}}$ 
above -6.5. 
Stronger correlations have been reported between
the non-thermal velocity and intensity of the 
\ion{Si}{4} line for the Quiet Sun by 
\citet{chae98}. Solar simulations suggests
that, depending on the magnetic field orientation,
the observed correlation between 
non-thermal broadening and line
intensity could be attributed to either 
shocks or turbulence \citep{depontieu15}. It
is however to be noted that these results
are based on spatially resolved
solar observations, which is not the 
same as the disk integrated stellar
observations.
Furthermore, unlike the Sun, where the correlation is 
for multiple observations of the 
same star, the 
correlation shown in Figure~\ref{fig8} 
is based on measurements taken for different
cool stars. The stellar correlation could be
due to different levels of magnetic activity, including 
short- and long-term variability. 

\subsection{Alfv\'en wave energy vs stellar rotation}\label{heating}
As discussed in Section \ref{energy2.4}, to 
determine the Alfv\'en wave energy density 
the mass density
must also be known in addition to 
the non-thermal velocity measurements.
We determined
the \ion{O}{4} 1399/1401 flux ratio to 
estimate the mass density $\rho$. 
{
The \ion{O}{4} lines are much weaker than 
the \ion{Si}{4}
lines (Figure~\ref{appendix_fig1}). Hence, the density estimates
could be obtained for only six stars out of 
our entire sample (Table~\ref{ratio}).
As show in Figure~\ref{appendix_ratio}, the 
line ratios lie close to the 
high-density limit of $N_\mathrm{e}=10^{12}$~cm$^{-3}$ 
in Figure~2 of \citet{polito16}. 
Hence, we 
applied the solar mass density \footnote {$N_\mathrm{e}$ of
2$\times$10$^{10}$~cm$^{-3}$ \citep{vanderholst14}}
in equation \ref{wave_eq} 
 as a minimum density to determine the Alfv\'en wave energy density
(Figure~\ref{appendix_fig10}). Since the 
density $\rho$ remains the same for all stars 
the 
trend is same as the non-thermal velocity.
A detailed density determination
will help to shed light into the dependence
of wave energy density on stellar properties.
}
\begin{table}
\caption{\label{ratio}Ratio of the \ion{O}{4}
lines at 1399 and 1401~\AA~for 
the sub-sample of stars in Figure~\ref{fig9}. Mean values
are shown for stars with multiple observations.}
\centering
\begin{tabular}{cc}
\hline
\hline
name&\ion{O}{4} 1399/1401\\
\hline
HD 22049&0.45\\
HD 75732&0.43\\
HD 72905&0.58\\
HD 1835&0.53\\
HD 201091&0.45\\
HD 39587&0.56\\
\hline
\end{tabular}
\end{table}

\subsection{Non-thermal velocity vs emission line temperature}
\subsubsection{The Sun}
\begin{figure}
\includegraphics[width=.5\textwidth]{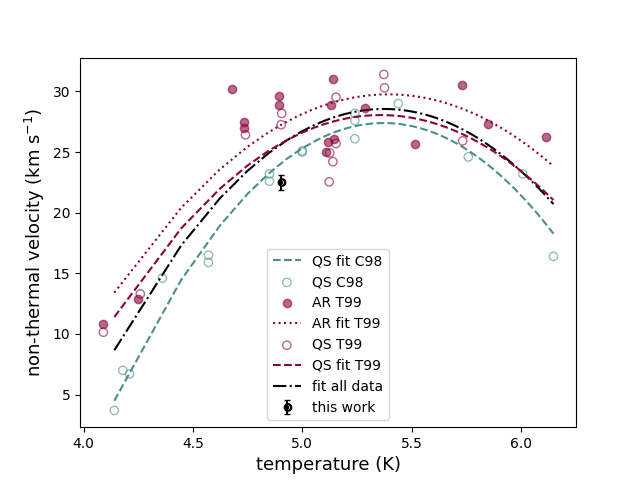}
\caption{Solar non-thermal velocities vs emission 
line temperature in log scale. The open circles show 
the quiet Sun (QS) measurements taken
from \citet{chae98} (green, C98) and \citet{teriaca99} (purple, T99), 
and the filled circles represent {active region
(AR)} measurements taken from \citet{teriaca99} (purple, T99). A second order
polynomial fit to the { AR} and QS measurements of T99 
is shown by the purple dotted and dashed line, respectively, while the dashed
green line marks a second-order polynomial fit to the QS measurements
of C98. The black dashed and dotted line is the overall fit obtained for 
all measurements included in this Figure (QS C98, QS T99, and { AR} T99). 
The mean Sun-as-a-star non-thermal velocity 
determined in this work is marked by the black $\odot$ with the 
standard deviation as error.
}
\label{arx2}
\end{figure}

The dependence of the solar non-thermal velocity
on emission line temperature in spatially resolved solar
measurements was
extensively studied in the past \citep{doschek76,chae98,teriaca99}. 
In Figure~\ref{arx2} we include 
non-thermal velocities, determined using 
SUMER observations, from two such studies
\citep{chae98,teriaca99}
and compare them to our Sun-as-a-star result
determined from the \ion{Si}{4} 1393 \AA~line. 
To make comparison easier we also fit a
second-order polynomial to the archival 
measurements, where green marks the fit obtained 
for \citet{chae98}, 
purple marks the fit obtained for \citet{teriaca99},
and black is the fit obtained for combined 
\citet{chae98} and \citet{teriaca99} measurements.
Table~\ref{polynomial} lists the equations of the 
polynomial fits.

\begin{table}
\caption{\label{polynomial}coefficients of the second
degree polynomial ({ $a \log^2 x+b\log x+c$})
fits in Figure~\ref{arx2}.}
\centering
\begin{tabular}{cccc}
\hline
\hline
&a&b&c\\
\hline
QS fit C98&-15.1&162.2&-408.2\\
AR fit T99&-10.39&112.1&-272.6\\
QS fit T99&-11.21&120.2&-293.9\\
fit all data&-13.06&140.3&-348.6\\
\hline
\end{tabular}
\end{table}

As shown in Figure~\ref{arx2}, the solar
non-thermal velocity increases with line temperature, 
peaks at around 100,000-200,000 K, and decreases 
towards higher temperatures. Figure~\ref{arx2} 
also shows that the {AR} measurements 
are consistently higher than the QS measurements.
The QS non-thermal velocities from \citet{chae98}
are weaker than the ones obtained by 
\citet{teriaca99}, which could be due to different observing
periods, and data reduction and analysis techniques. 
Despite this slight discrepancy the overall
trend exhibited by both of these studies is very similar.
Our Sun-as-star non-thermal velocity is on 
the lower end, 
closest to  
the QS value of \citet{chae98}.
\begin{figure*}
\centering
\includegraphics[width=0.9\textwidth]{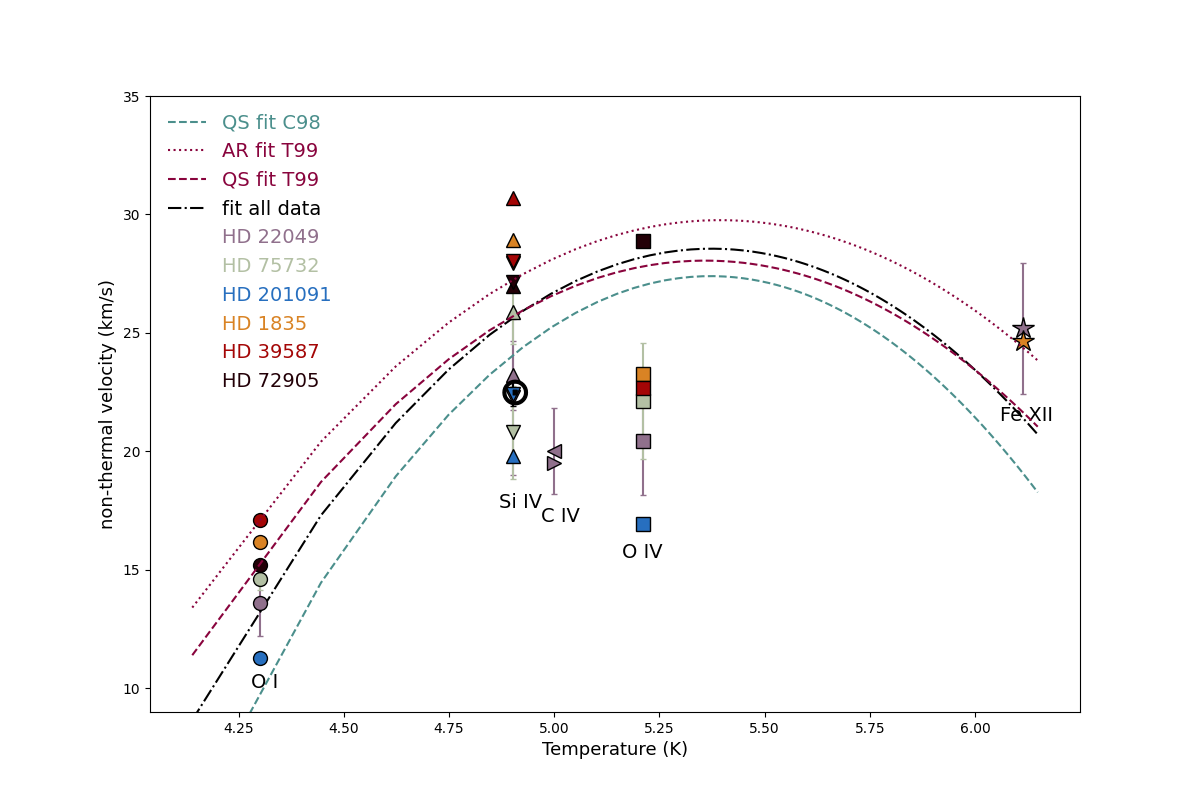}
\caption{Non-thermal velocity vs emission line temperature. Individual 
stars are represented by the colour and the spectral lines
are represented by the symbols. 
The error bars represent the dispersion for stars with multiple 
measurements. Each star is colour coded as shown in labels.The mean solar non-thermal
velocity determined from the \ion{Si}{4} 1393~\AA~line is shown 
by the solar symbol. The coloured lines represent the polynomial 
fits from Figure~\ref{arx2}.}
\label{fig9}
\end{figure*}
\subsubsection{Sun-like stars}
To determine the dependence of 
non-thermal velocity
on emission
line temperature in stars other than the Sun we analysed
the seven emission lines discussed in 
Section~\ref{lines} and determined the non-thermal
broadening for six stars with appropriate
wavelength coverage and good 
signal-to-noise ratios. Although six is a small number the stars vary in 
mass, age and rotation period, and
include 
two older Sun-like stars 55 Cnc and 
61 Cyg A, and four active young
Sun-like stars $\pi^1$ UMa, $\epsilon$ Eri, 
9 Cet and $\chi^1$Ori.
Two out of these six stars are known
exoplanet hosts, 55 Cnc and $\epsilon$ Eri.
Out of these six stars 
the non-thermal broadening in the
\ion{C}{4} doublet could be determined for only one
target, $\epsilon$ Eri, as this line 
was out of the wavelength coverage for the 
rest of the stars. Additionally, due to 
poor signal-to-noise ratio of the 
\ion{Fe}{12} line we could
determine the non-thermal
broadening in this line for
two out of the six stars, $\epsilon$ Eri 
and 9 Cet. The 
\ion{O}{1}, \ion{Si}{4}, and the \ion{O}{4}
lines could be analysed for all
six stars.

Figure~\ref{fig9} shows the non-thermal broadening
vs emission line temperature 
for the six stars discussed above, where
each star is represented by a single colour.
Similar to the solar case, the 
stellar non-thermal velocities 
exhibit a clear dependence
on emission line temperature.
The non-thermal velocity is the lowest in the 
chromosphere (20000 K)
and the highest at the transition region. However,
unlike the Sun where the peak occurs between 
100,000-200,000 K,
in almost all stars the non-thermal
velocity peaks at a lower temperature of 80,000 K.
HD72905 or $\pi^1$ UMa is the only star analysed in this work
where
the non-thermal velocity peak is similar to the Sun.
The other young solar analogues in the sample 
do not exhibit the same trend.
A similar analysis by \citet{pagano04} on HST
STIS spectra of the moderately 
active star $\alpha$ Cen A 
showed that its non-thermal
velocity peaks at a similar temperature to the Sun.
This suggests that the discrepancy between the
solar and stellar non-thermal velocity peaks, seen 
in Figure~\ref{fig9}, might not be related to 
spatial scales and instrument used, 
but could be due to differences in the underlying 
physical processes that causes the 
non-thermal broadening in these lines.

As discussed previously, the non-thermal broadening
of emission lines is due to mechanisms such as, Alfv\'en
waves, turbulence, shocks, flares, and re-connection events. 
It is now widely accepted that Alfv\'en waves are the 
dominant transport mechanism in the Sun. 
Based on the relationship between
the non-thermal velocity and emission line temperature in
$\pi^1$ UMa and $\alpha$ Cen A, it is reasonable to 
assume that the energy transport processes 
in these two stars are very similar to the Sun.
This provides a strong justification to 
the suitability of
Alfv\'en wave driven models for wind simulations
in stars like $\pi^1$ UMa and $\alpha$ Cen A.

For some stars, however, the non-thermal 
velocity peak at~80,000 K indicates that
a majority of the 
Alfv\'en waves are either dissipating at 
lower temperatures than the 
Sun, or processes
other than Alfv\'en waves are contributing
towards the non-thermal
broadening for the \ion{Si}{4} line
at 80,000 K.
As an example, the $\epsilon$ Eri
observations used in this work are 
taken from the HST archive and some of these
observations are impacted by flares
\citep{loyd22}. It is however difficult 
to judge how extreme events such as
flares impact the non-thermal broadening of
the interface region emission lines from these
observations alone. A detailed study of these 
emission lines at different levels of solar activity,
including and excluding large flare events could help
shed further light into this problem.

\section{Summary and conclusion}
We analysed HST COS and IRIS archival spectra to 
determine the non-thermal velocity of cool
Sun-like stars by measuring the non-thermal broadening
in the \ion{Si}{4} line for 56 stars including the Sun.

To determine the non-thermal velocity of the Sun 
we used full-disc mosaics of IRIS, which can be compared
to disk-integrated stellar observations. The non-thermal
velocities obtained this work
are in good agreement with results obtained from 
spatially resolved observations of the quiet Sun and also
with other Sun-like stars that have similar spectral type 
and rotation period. We also
detect a weak correlation between the solar 
non-thermal velocity and the sunspot numbers. 

The \ion{Si}{4} 1393~\AA~ line profile 
was also used to determine the
non-thermal velocity of 55 other Sun-like stars, 
the majority of
which are exoplanet hosts. The non-thermal velocity
shows a clear dependence on rotation, where an increase
in rotation rate is followed by an increase in non-thermal velocity.
We also determined the Alfv\'en wave energy density, assuming that the
non-thermal broadening is caused by the { presence} of transverse
Alfv\'en waves. {By applying the solar plasma 
density in the transition region as a representative
minimum value for cool stars, our
}results show that the Alfv\'en wave energy 
density follows the same trend as the non-thermal velocity. 
Since the Alfv\'en wave energy density
is an important input parameter for state-of-the-art 
stellar wind models such as AWSoM \citep{vanderholst14}, these results provide, 
for the first time, a data-driven way to constrain the 
wave energy in such models, and thus help us scale the models for 
stars with different rotation periods. {However, detailed 
density determination in cool stars is still needed to fully
exploit the diagnostic capabilities of the emission lines discussed here}.

Finally, we investigated the relationship between 
non-thermal broadening and emission line temperature 
using seven emission lines with temperatures ranging from 
from the chromosphere to the lower corona.This part of the analysis could be only 
applied to a limited
sample size of six stars due to signal-to-noise ratio constraints. 
In the Sun the non-thermal velocity 
increases with temperature, peaking at around 100,000-200,000 K and
decreases towards coronal temperatures. Our results 
show that in cool stars the global trend for
non-thermal velocity vs emission line temperature is very similar to 
the Sun, however the non-thermal velocity peaks at a 
lower temperature for some stars. This suggests that,
in some cool stars,
either a majority of the Alfv\'en energy is dissipated 
at a lower temperature in the transition region, 
or other processes, such as flares could contribute 
strongly towards the non-thermal broadening at low
temperatures. To obtain a better understanding of 
the contribution of Alfv\'en  waves and energetic 
events such as flares, detailed investigations
of the Sun and stars using emission lines formed at
different interface region temperatures is required.
\begin{acknowledgements}
This research was funded by the Austrian Science Fund (FWF)
Lise Meitner grant: M2829-N. V.S.A. acknowledges support from 
the NASA/GSFC Sellers Exoplanet Environments
Collaboration (SEEC), which is funded by the 
NASA Planetary Science Division’s Internal
Scientist Funding Model (ISFM) and funding from 
HST GO Cycle 27 NAS5-26555. MJ is supported by NASA's SDO/AIA 
contract (NNG04EA00C) to LMSAL. PT was supported by contract 
8100002705 (IRIS) from Lockheed-Martin to SAO.
\end{acknowledgements}
\bibliography{ref}{}
\bibliographystyle{aasjournal}
\appendix
\restartappendixnumbering
\renewcommand{\theHfigure}{A\arabic{figure}}\begin{figure*}
\section{Absolute flux densities}
\centering
\includegraphics[width=1\textwidth]{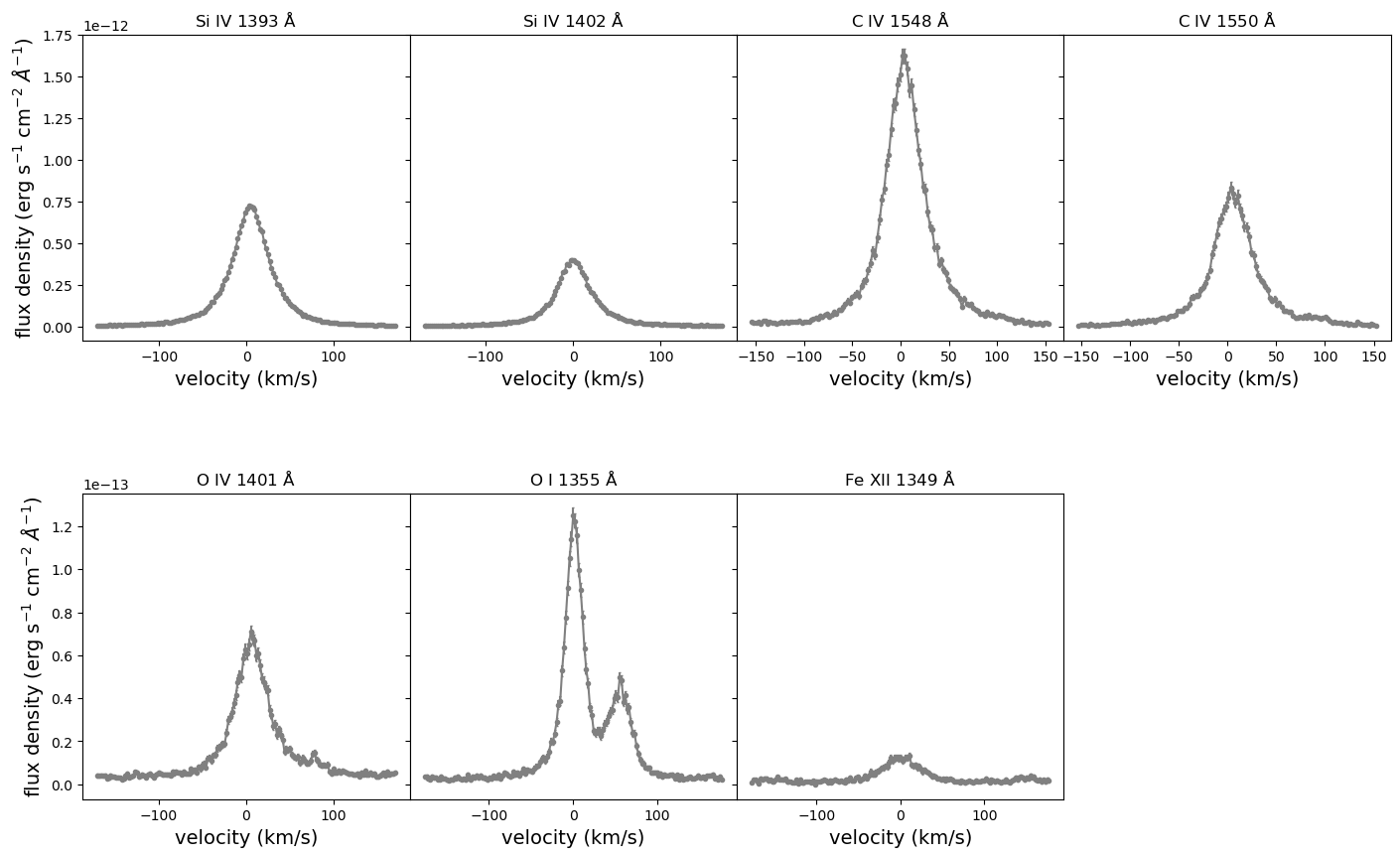}
\caption{Same as Figures~\ref{fig1} and \ref{fig2}, but absolute flux densities
are shown instead of the normalised values. For better visualisation
the y-axis is shared row-wise.}
\label{appendix_fig1}
\end{figure*}

\restartappendixnumbering
\renewcommand{\theHfigure}{A\arabic{figure}}

\begin{figure}
\section{Alfv\'en wave energy density}
\centering
\includegraphics[width=0.5\textwidth]
{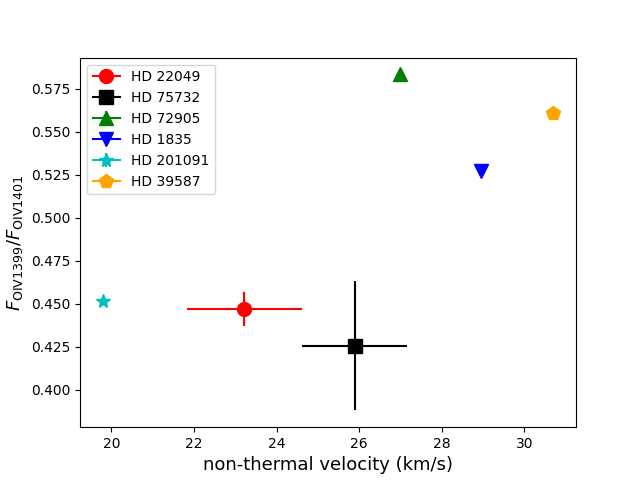}
\caption{\ion{O}{4} flux ratio vs non-thermal velocity. The coloured symbols
mark different stars. For stars with multiple measurements the mean is plotted
with the standard deviations as errorbars.}
\label{appendix_ratio}
\end{figure}

\begin{figure}
\centering
\includegraphics[width=0.5\textwidth]
{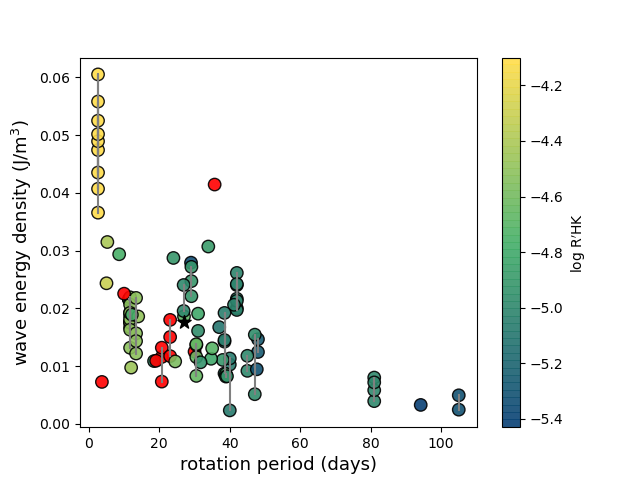}
\caption{Wave energy density 
vs rotation period. Colour scale same as Fig~\ref{fig5}.}
\label{appendix_fig10}
\end{figure}

\restartappendixnumbering
\renewcommand{\theHfigure}{A\arabic{figure}}\section{IRIS observations included in A21}\label{comparison}
\begin{figure}
\centering
\includegraphics[width=.7\textwidth]{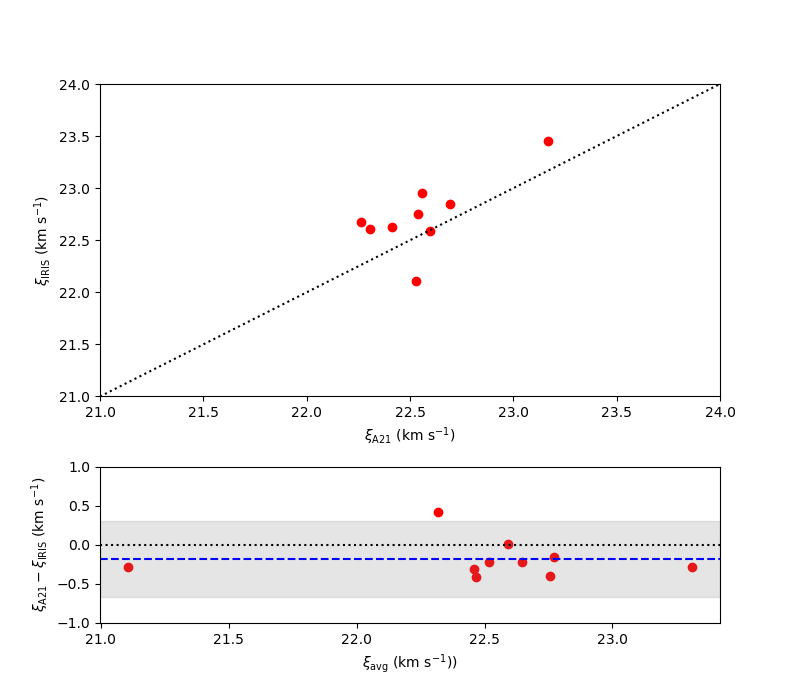}
\caption{Non-thermal velocities $\xi$ for the 10 observations in
Table~\ref{comparison_table}. {\it Top}: The x-axis represents
measurements on data-set A and the
y-axis represents measurements carried out on data-set B. The 
dashed black line is the line of equality. {\it Bottom}: Mean-difference
plot, where the x-axis represents the mean and the y-axis shows 
the difference between data-set A and B. The 
dotted black line marks where the difference is zero. The dashed
blue line marks the bias or average difference, and the grey
area marks the bias+2$\sigma$.}
\label{fig_new}
\end{figure}

In order to determine how individual 
observations are impacted by the data
processing step of A21 we compared the 
A21 observations (data-set A) with IRIS observations
that didn't go through the additional 
data-processing steps of A21 (data-set B). 
The non-thermal velocities obtained from 
these two data-sets are listed 
in Table~\ref{comparison_table}.
Figure~\ref{fig_new} shows a comparison 
plot between these two data-sets, including a mean-difference plot. 
The top plot in Figure~\ref{fig_new} shows that 
all except one measurement in
data-set B are slightly higher 
than the values obtained from data-set A. 
Overall, no significant difference is detected
between the two data-sets, as shown by the 
bottom plot in 
Figure~\ref{fig_new} where the bias or the average 
difference between the 
two data-sets is close to zero and all except one 
measurement falls inside the limits of agreement (bias $\pm2\sigma$).

\begin{table}
\centering
\caption{\label{comparison_table} Solar 
non-thermal velocities. Columns 1 to 3: date or observation
ID, and
results for data-set A ($\xi_\mathrm{A21}$) and data-set B ($\xi_\mathrm{IRIS}$) respectively. }
\begin{tabular}{ccc}
\hline
\hline
date&$\xi_\mathrm{A21}$&$\xi_\mathrm{IRIS}$\\
year/month/day&km~s$^{-1}$&km~s$^{-1}$\\
\hline
2013/10/27& 21.1& 21.4\\
2014/03/17& 22.4& 22.9\\
2014/03/24& 22.5& 22.9\\
2015/04/01& 23.3& 23.7\\
2015/10/18& 22.7& 23.2\\
2017/10/21& 22.8& 23.1\\
2018/08/25& 22.7& 23.0\\
2019/04/13& 22.4& 22.9\\
2019/05/05& 22.7& 22.8\\
2019/09/12& 22.6& 22.4\\
\hline
\end{tabular}
\end{table}
\restartappendixnumbering
\renewcommand{\theHfigure}{A\arabic{figure}}

\end{document}